\documentclass{article}
\usepackage{spconf,amsmath,graphicx}
\usepackage{multirow}
\usepackage{booktabs}
\usepackage{amssymb}
\usepackage{bm}
\usepackage{float}
\usepackage[export]{adjustbox}

\usepackage{enumitem}
\setlist{nosep, leftmargin=14pt}
\usepackage[left=1.62cm,right=1.62cm,top=1.9cm]{geometry}

\title{Causal Modeling of fMRI Time-series for Interpretable Autism Spectrum Disorder Classification\\ }

\address{
$^{1}$ Department of Biomedical Engineering, Yale University, USA\\
$^{2}$ Radiology \& Biomedical Imaging, Yale School of Medicine, USA\\
$^{3}$  Electrical Engineering, Yale University, USA\\}
\begin{document}
%
\maketitle

\begin{abstract}
Autism spectrum disorder~(ASD) is a neurological and developmental disorder that affects social and communicative behaviors. It emerges in early life and is generally associated with lifelong disabilities. Thus, accurate and early diagnosis could facilitate treatment outcomes for those with ASD. Functional magnetic resonance imaging (fMRI) is a useful tool that measures changes in brain signaling to facilitate our understanding of ASD. Much effort is being made to identify ASD biomarkers using various connectome-based machine learning and deep learning classifiers. However, correlation-based models cannot capture the non-linear interactions between brain regions. To solve this problem, we introduce a causality-inspired deep learning model that uses time-series information from fMRI and captures causality among ROIs useful for ASD classification. The model is compared with other baseline and state-of-the-art models with 5-fold cross-validation on the ABIDE dataset. We filtered the dataset by choosing all the images with mean FD less than 15mm to ensure data quality. Our proposed model achieved the highest average classification accuracy of 71.9\% and an average AUC of 75.8\%. Moreover, the inter-ROI causality interpretation of the model suggests that the left precuneus, right precuneus, and cerebellum are placed in the top 10 ROIs in inter-ROI causality among the ASD population. In contrast, these ROIs are not ranked in the top 10 in the control population. We have validated our findings with the literature and found that abnormalities in these ROIs are often associated with ASD. 

\end{abstract}
\section{Introduction}
\label{sec:intro}
\let\thefootnote\relax\footnotetext{\textcopyright 2025 IEEE.  Personal use of this material is permitted.  Permission from IEEE must be obtained for all other uses, in any current or future media, including reprinting/republishing this material for advertising or promotional purposes, creating new collective works, for resale or redistribution to servers or lists, or reuse of any copyrighted component of this work in other works.}
The human brain is a complex structure and operates as a constantly communicating and dynamic network. To study the change in brain activities, functional magnetic resonance imaging~(fMRI) is often used to measure the blood-oxygen-level-dependent~(BOLD) signals, which could provide insight into the diagnosis and treatment of brain disorders such as autism spectrum disorder~(ASD) \cite{fmri-asd}. fMRI could also be utilized to study the connectivity between brain regions, revealing abnormal connectivity patterns associated with ASD.

A functional connectivity (FC) map is often reconstructed by calculating the correlations between fMRI signals of regions of interest~(ROIs) to represent the brain networks. In recent years, machine learning models, such as CNNs, RNNs, and GNNs, have been developed to identify abnormal patterns in FC maps to help classify ASD \cite{braingnn}\cite{stagin}\cite{spectbgnn}\cite{brainnet_cnn}\cite{fmri-pretarin-transformer}. However, FC is inherently linear, while we know that interactions in functional networks are complex and nonlinear \cite{correlation-limitation}\cite{science-brain}. Furthermore, the correlation between brain regions does not mean causality. Thus, developing techniques that leverage the temporal structure of BOLD signals to identify brain region interactions is crucial to investigating brain mechanisms. 

One approach for inferring causality among time series is based on the Wiener-Granger principle \cite{granger-causality}, which could measure directed relationships between multiple ROIs of the brain by fitting a vector autoregressive model to forecast time series \cite{var_paper}. In this paper, we propose a new causal neural network to identify subjects with ASD with Granger-inspired interpretability by expanding the concept of multivariate vector autoregressive (VAR) \cite{granger-causality} \cite{var_paper}. Moreover, the frequency loss function is used to train the network for better performance \cite{fredf}. We further compare and validate our model performance with state-of-the-art models with a 5-fold cross-validation on the filtered Autism Brain Imaging Data Exchange I (ABIDE I) dataset \cite{abide}. Our model captures and explains the top 10 highest Granger-inspired causal ROIs. 

\section{Method}
In this paper, we propose a new supervised learning network that takes ROI-level BOLD fMRI signals as input and outputs ROI-level BOLD fMRI signal forecasts and ASD classification results. The causality encoder is a single-layer long-short-term memory (LSTM). The encoded causality embeddings are further processed by the attentional pooling layer and a fully connected neural network for classification. The causality decoder is a 2-layer LSTM that decodes the causality embeddings. The decoder output could be compared with the real ROI time series to measure predictability. 

\begin{figure*}[!t]
\centering
 \includegraphics[clip, trim=0.5cm 5cm 0cm 2.5cm, width=\linewidth]{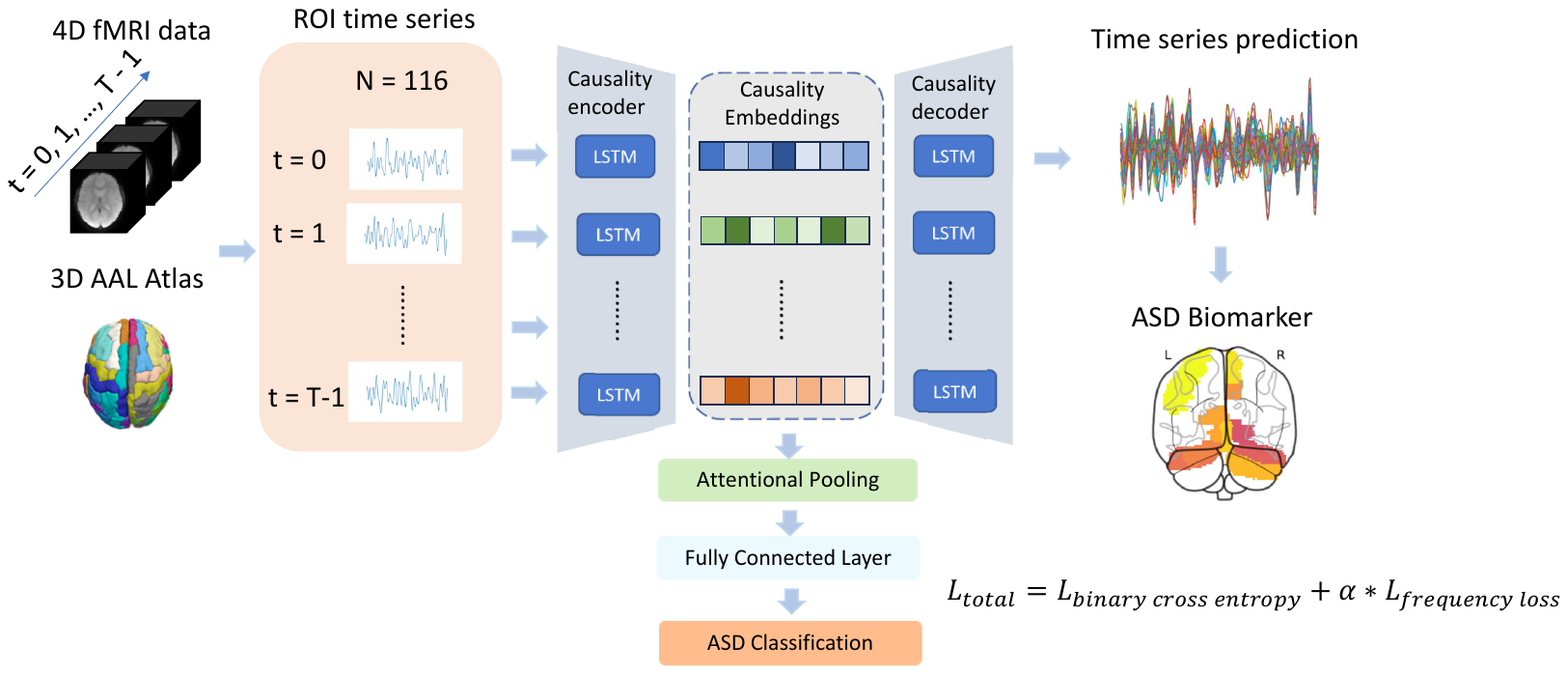}
    \caption{Architecture of the proposed network. The 4D fMRI data is combined with the 3D AAL atlas and the fMRI signals in the same ROI are averaged for each time point. The multichannel ROI time series from t=0 to t=T-1 is encoded by the causality encoder made up of long-short-term memory (LSTM) modules, which are used to both (1) predict future time points and (2) predict ASD classification labels. The attentional pooling layer further processed the causality of multichannel embeddings. The fully connected layer then made an ASD classification prediction. The causality decoder decodes causality embeddings and predicts the future time point from t=1 to t=T.}
  \label{model_architecture}
\end{figure*}

\subsection{Causal inspired fMRI ROI time series forecasting}
We define fMRI ROI time series as $X$ and each ROI time series $X_{i}$ where $i = \{1, 2, ..., N\}$ represents each ROI in the brain, and $N$ is 116 determined by the total number of regions in the Automated Anatomical Labeling (AAL) atlas \cite{aal_atlas}. $X_{i}(t)$ is the signal at time $t$ where $t = 0, \Delta T, ..., T$, $\Delta T$ is the sampling rate and $T$ is the total duration of the acquisition. In multivariate scenarios, the Granger causality could be captured by the linear VAR model $X(t) = \sum^L_{\tau=1} A_{\tau}X(t-\tau) + \epsilon(t)$, where $\tau$ is the predetermined time lag, L is the total time lag and $\epsilon$ is the random error \cite{var_paper}. $A_{\tau}$ is a matrix for every different $\tau$ for the mixing of different variables. The limitation of this model is the linear relationship between past and future time points, which could be an oversimplification of brain data, while deep learning models could capture more complex causal relationships. 

We create input and target sequences separated by a time lag of $\tau$ seconds to train the model to forecast future ROI time series.  Specifically, we discard the initial $\tau$ seconds of the input to form the target sequence, while the last $\tau$ seconds of the fMRI data are removed to create the input sequence. The structure of our network is illustrated in Fig.~\ref{model_architecture}. Our model takes all the previous time points and predicts the next. The time series forecast can be expressed as: 
\begin{equation}
\phi\big(X_{1}(t - \tau), X_{2}(t - \tau), \dots, X_{N}(t - \tau)\big) \approx X_{i}(t)
\label{eq:1}
\end{equation}
\begin{center}
for \( t = 1, 2, \ldots, T - \tau,\ i=1,2,\ldots,N\)
\end{center}
$\phi$ is the non-linear function learned by our encoder-decoder path of the model. The choice of $\tau$ is 1 TR because the fMRI sample rate is low \cite{timelag}. 

After training, we evaluate the model based on inter-ROI causalities inspired by Granger causality. Different from the original VAR model where the causality between all ROIs is expressed in the vector $A_{\tau}$, the neural network could not be simply expressed in the format of linear multiplication. However, we could measure how close the time series forecast is to the real-time series. The inter-ROI causality depends on the function in the class $F=\{f(X,\hat{X}):f:\mathbb{X}\times \mathbb{X}\rightarrow \mathbb{Z}\}$, where $f$ measures the similarity of the true time series $X$, and the predicted time series $\hat{X}$ based on the previous time points. The mean squared error function could help as a proxy for measuring predictability as lower values mean better similarity between true time series and forecast generated by the model. The predictability can be measured by:
\begin{align*}
\mathcal{P}_{i} &= 1 - \frac{\text{E}}{\text{Var}(X_{i})} \\
\text{E} &= \frac{1}{n} \sum_{i=1}^{n} \left(X_{i} - \hat{X}_i \right)^2 \\
           &= \frac{1}{n} \sum_{i=1}^{n} \left( X_{i}(t) - \phi\big(X_{1}(t - \tau),  \dots, 
           X_{N}(t - \tau)\big) \right)^2 \\
           &\quad \quad t = 1, 2, \dots, T - \tau
\end{align*}

Here, $\mathcal{P}$ represents predictabilty. As error E approaches 0, $\mathcal{P}$ approaches 1, indicating perfect predictability. 

\subsection{Frequency loss function}
In previous studies, frequency domain information of ROI time series has proven useful for ASD label prediction \cite{spectbgnn}. In this study, we train the model with the Frequency Loss function which captures the ROI fMRI signals frequency domain features \cite{fredf}. As shown in Equation \ref{eq:frequency_loss}, the frequency loss is calculated based on the differences in the discrete Fourier transforms. The absolute value is used instead of the square to avoid overemphasizing the frequency bandwidth with a large spectrum. 
\begin{equation}
L_{\text{freq}} = \sum_{k=0}^{K-1} | \text{DFT}(\hat{X})(k) - \text{DFT}(X)(k) |
\label{eq:frequency_loss}
\end{equation}

\subsection{ASD classification with Attentional Pooling}
We use an attentional pooling module to compress the multichannel causal embedding before feeding it into the fully connected layer to generate the ASD classification. The multihead attention layer first projects the embedding with 3 linear layers into 3 vectors: $Q$, $K$, and $V$ \cite{attention}. Attention is calculated as shown in Equation \ref{eq:attention}. $d_k$ is the dimension of the vector $K$. By dividing the dot product $QK$ by $\sqrt{d_k}$, we could control the magnitude of $QK$ to avoid network training problems such as gradient explosion. We use multiple sets of $Q$, $K$, and $V$ so that the model can focus on multiple aspects of the embedding. Finally, a linear layer combines these learned outputs into the original size as the embedding. We use the validation dataset to experiment with the number of heads in our application. 
\begin{equation}
{\text{Attention}}(Q,K,V) = \text{softmax}(\frac{QK^{T}}{\sqrt{d_k}})V
\label{eq:attention}
\end{equation}

The embedded representation refined by the pooling layer is then fed into a dense network for the ASD classification. Binary Cross Entropy loss is combined with the frequency loss function to train the network by scaling the frequency loss function as shown in Equation \ref{total_loss}. $y_i$ denotes the actual ASD/control label of the i-th subject, $S$ is the total number of subjects and $p_i$ is the predicted possibility of ASD. 
\begin{equation}
L_{total} = \sum_{i=1}^{S}-{(y_i\log(p_i) + (1 - y_i)\log(1 - p_i))} + \alpha L_{freq(i)}
\label{total_loss}
\end{equation}

\section{Experiments}
\label{sec:experiments}
\subsection{Datasets and pre-processing}
\textbf{ABIDE I} The resting-state fMRI data was obtained from the ABIDE I database \cite{abide}. ABIDE I is collected from 17 international sites, and the neuroimaging and phenotypic data of 1112 subjects are publicly shared. As a quality control, fMRI images with mean FD larger than 0.15mm are excluded from this study. The total number of subjects filtered is 860. 

\noindent\textbf{Implementation Details}~~~~ We set the learning rate to train the network to 0.005 and run 100 epochs. The learning rate is reduced to half for every 8 epochs. We apply the frequency loss function, and binary cross entropy loss as the loss function and Adam Optimizer to train the network \cite{adam}. 

\noindent\textbf{Baselines and Ablation Studies}~~~~ We have compared our proposed method with various baseline methods. One baseline method, which does not involve deep learning, is Connectome-based Predictive Modeling (CPM) using ridge regression \cite{cpm-short}. The hyperparameter alpha in CPM, representing the coefficient of the regularization term, is searched within the range of 0.5 to 5x$10^9$. The optimal alpha value is selected by assessing the model's performance on a validation dataset.
We also compare the model with the graph-based methods STAGIN \cite{stagin}, BrainGNN \cite{braingnn}, and SpectBGNN \cite{spectbgnn} and an attention-based approach BolT \cite{bolt}. \\
Furthermore, we also test the choice of $\alpha$, and the trade-off between training the network for better time series forecasts or better ASD label predictions. We have experimented with different scales of $\alpha$ from 0.1 to 0.0001 to determine the best $\alpha$ to use for training the network. The $\alpha$ we would like to select is the one that yields high accuracy and comparable frequency loss in the validation set. 

\noindent\textbf{Evaluation Methods}~~~~
We use 5-fold cross-validation by randomly splitting the dataset into train, test, and validation datasets. We apply each model and measure the accuracy, AUC, recall, and precision to compare the performance. We measure the mean absolute error to test the best suitable number of layers of the LSTM decoder. 

\section{Results}
\begin{table*}[h!]
\caption{The model performance in 5-fold cross-validation is summarized. The bolded has the highest average of all models.}
\centering
\begin{tabular}{c|c|c|c|c}
\hline
\textbf{Model} 
&\textbf{Accuracy $\uparrow$} 
& \textbf{AUC $\uparrow$}
& \textbf{Recall $\uparrow$}
& \textbf{Precision $\uparrow$}
\\ 
\hline
\text{CPM} & $59.4\% \pm 5.2\%$ & $66.7\% \pm 2.7\%$ & $53.5\% \pm 10.5\%$ & $56.1\% \pm 5.4\%$\\
\text{BrainGNN} & $61.6\% \pm 1.0\%$ & $60.4\% \pm 4.0\%$& $61.3\% \pm 11.5\%$ & $48.1\% \pm 3.5\%$\\
\text{SpectBGNN} & $61.5\% \pm 0.5\%$ & $60.0\% \pm 1.6\%$ & $63.4\% \pm 9.3\%$ & $45.3\% \pm 10.5\%$\\
\text{STAGIN} & \textbf{$69.4\% \pm 0.8\%$} & \textbf{$75.6\% \pm 0.7\%$} & $\mathbf{74.2\% \pm 1.3\%}$ & \textbf{$71.7\% \pm 1.7\%$}\\
\text{BolT} & \textbf{$68.0\% \pm 2.7\%$} & \textbf{$74.3\% \pm 3.7\%$} & \textbf{$73.5\% \pm 4.5\%$} & \textbf{$68.6\% \pm 2.4\%$}\\
\hline
\textbf{Ours} & $\mathbf{71.9\% \pm 0.8\%}$ & $\mathbf{75.8\% \pm 1.0\%}$ & \textbf{$59.6\% \pm 8.9\%$} & $\mathbf{76.5\% \pm 11.7\%}$\\
\hline
\end{tabular}

\label{tab:accuracy_results}
\end{table*}

\subsection{Experimental Results} 

\noindent\textbf{ASD classification performance}~~~~ All of the baseline methods performance are shown in Table \ref{tab:accuracy_results}. Deep learning-based methods have better average accuracy and less variance than CPM. Our proposed method shows the best result in the accuracy AUC, and precision than the rest shown in Table \ref{tab:accuracy_results}.

\noindent\textbf{Choice of }$\mathbf{\alpha}$~~~~
Fig.~\ref{alpha_effect} shows the classification performance measured by accuracy and time series prediction performance measured by scaled frequency loss with the validation set. Overall, the lower the $\alpha$, the better performance in the classification of ASD subjects. However, when $\alpha$ is lower than $1 * 10^{-4}$, the accuracy does not increase. Therefore, our best model is trained with $\alpha=1* 10^{-4}$.

\noindent\adjustimage{width=.45\textwidth,center,caption={Multiple $\alpha$ values performance are visualized. For ease of interpretation, the frequency loss is scaled by dividing the maximum frequency loss in the experiment.},label={alpha_effect}, figure}{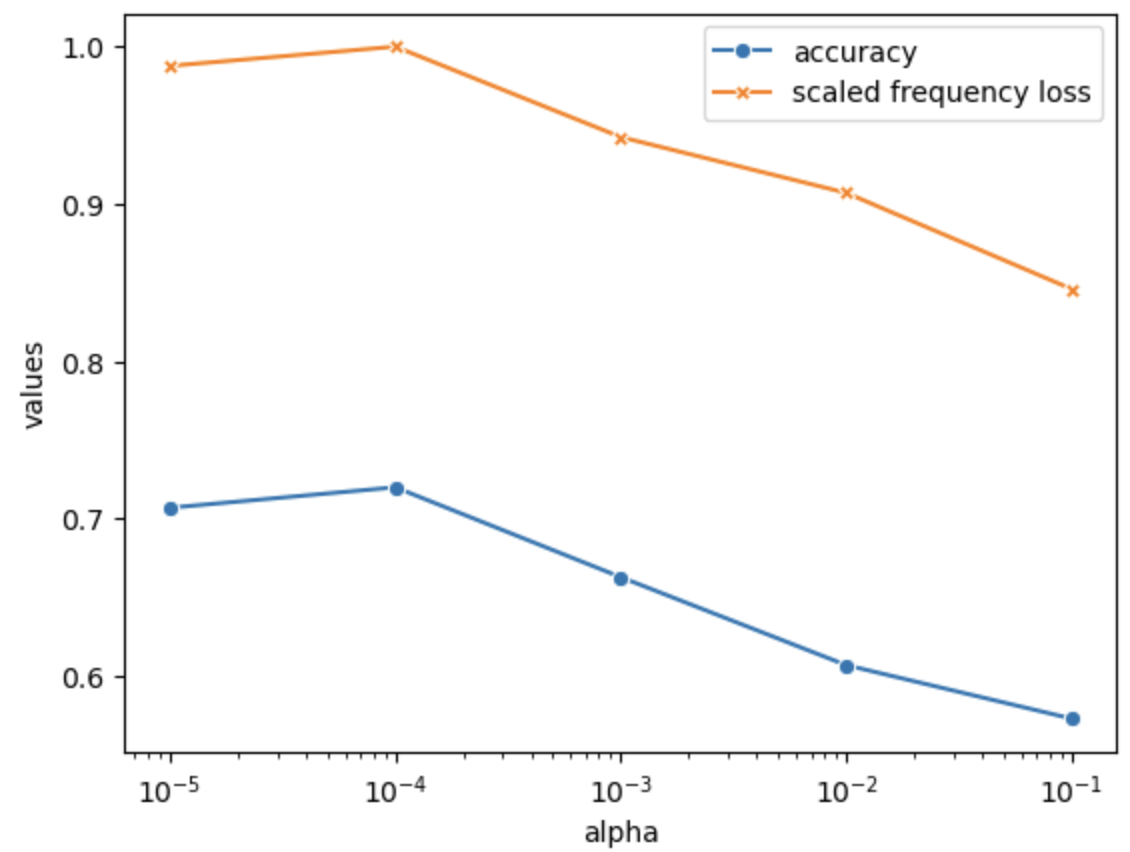}

\noindent\textbf{Predictability rank}~~~~
We ranked each subject's ROI by their predictability defined in the previous section. We could obtain the occurrence of each ROI that is ranked in the top 10 predictable ROIs from each individual whom our model correctly classified as ASD or control. Then we could get the top 10 most frequent ROI occurrences for both control and ASD populations. Compared to the control population, the population with ASD has the left precuneus, right precuneus, and left cerebellum lobule IV/V selected for the top 10 ROIs (shown in Fig.~\ref{significant_ROI}). The precuneus is associated with communication and recent studies have found that it has different connectivity patterns in the ASD population than control groups \cite{precuneus}. The cerebellum is also reported to be connected to other social brain areas and shown to be significantly associated with ASD pathology \cite{cerebellum}. The control population has left precentral gyrus, right supplementary motor area, and left calcarine fissure in the top 10 that the ASD population does not have. The top 10 ASD predictability ranges from 0.09 to 0.2, while the top 10 ROI predictability ranges from 0.07 to 0.2. Both ASD and control populations have these ROIs in common: right mid-cingulum, left cerebellum, vermis, right lingual gyrus, left cerebellum crus I, right cerebellum crus I, and right cerebellum crus II. This means that these ROIs are important to both populations in the resting state, but this could not be interpreted as a denial of the association between these ROIs and ASD.

\noindent\adjustimage{width=.48\textwidth, 
left,
caption={ROIs with high predictability for ASD and control population shown in MNI space.},label={significant_ROI},figure}{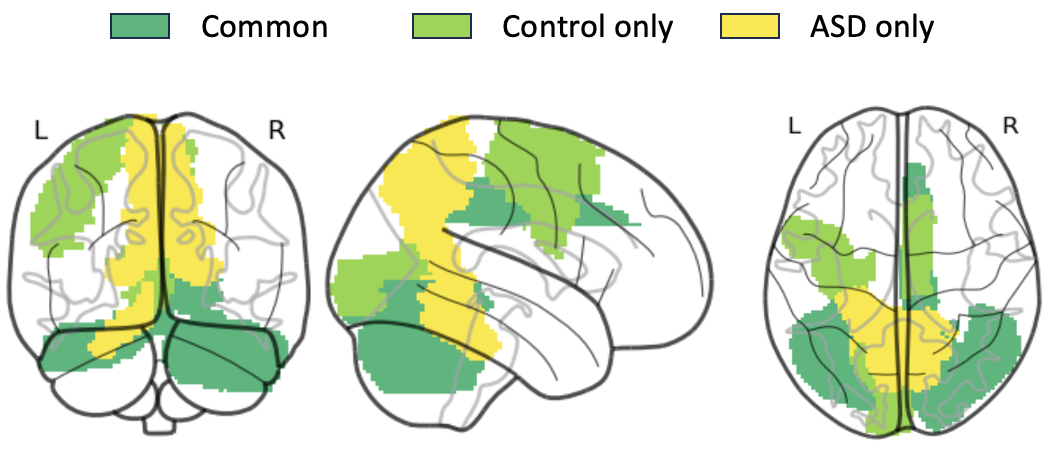}

\section{Discussion and Conclusions}
We propose a new network to classify ASD with the interpretability of causality. We use a combined frequency loss function and binary cross-entropy function to train the network to forecast time series and classify ASD simultaneously. The scaling factor to combine these functions is selected with thorough experiments. Our model is compared with other baseline methods and achieves a higher average ASD classification accuracy and AUC. 

Furthermore, we identify the left precuneus, right precuneus, and cerebellum to have a higher ranking in the inter-ROI causality relationship in the ASD population than the control. These ROIs are validated by other studies which reported these ROIs to be associated with ASD. 
 
\section{Compliance with Ethical Standards}
This research study was conducted retrospectively using human subject data made available in open access by NITRC IR. Ethical approval was not required as confirmed by the license attached with the open-access data. 
\section{Acknowledgments}
\label{sec:acknowledgments}

 The authors would like to thank all participants. This study is supported by the National Institute of Neurological Disorders and Stroke~(NINDS) of the National Institutes of Health through grant R01~NS035193.

\bibliographystyle{IEEEbib}
\bibliography{ISBI2025_ASD}

\end{document}